\documentclass[pdflatex,sn-aps]{sn-jnl}%
\jyear{2021}%

\theoremstyle{thmstyleone}%
\theoremstyle{thmstyletwo}%
\newcommand{\g}{\tilde{T}}
\newcommand{\s}{\mathbf{\hat{s}}(\phi)}
\newcommand{\n}{\mathbf{\hat{n}}}
\newcommand{\kn}{\mathrm{Kn}}

\theoremstyle{thmstylethree}%

\begin{document}

\title[Inverse Design in Nanoscale Heat Transport]{Inverse Design in Nanoscale Heat Transport via Interpolating Interfacial Phonon Transmission}

%Differentiable Phonon Simulations To Optimize
%Thermal Transport in Nanostructures}


\author*[1]{\fnm{Giuseppe} \sur{Romano}}\email{romanog@mit.edu}

\author[2]{\fnm{Steven} \sur{G. Johnson}}

\affil[1]{\orgdiv{Institute for Soldier Nanotechnologies}, \orgname{Massachusetts Institute of Technology}, \orgaddress{\street{77 Massachusetts Avenue}, \city{Cambridge}, \postcode{02139}, \state{MA}, \country{USA}}}

\affil[2]{\orgdiv{Department of Mathematics}, \orgname{Massachusetts Institute of Technology}, \orgaddress{\street{77 Massachusetts Avenue}, \city{Cambridge}, \postcode{02139}, \state{MA}, \country{USA}}}


%\abstract{We introduce a methodology for \textcolor{blue}{density-based topology optimization of non-Fourier thermal transport in nanostructures, based upon the forward and adjoint phonon Boltzmann transport equation (BTE)}. To this end, we also develop the transmission interpolation model (TIM), an interface-based method that allows for smooth interpolation between void and solid regions. We first use our approach to tailor the effective thermal conductivity tensor of a periodic nanomaterial; then, we maximize classical phonon size effects under constrained diffusive transport, \textcolor{blue}{identifying a promising thermoelectric material based on a structure with a relatively simple topology}. Our method enables systematic optimization of materials for heat management and conversion, and, more broadly, the design of devices where diffusive transport is not valid.}

\abstract{We introduce a methodology for density-based topology optimization of non-Fourier thermal transport in nanostructures, based upon adjoint-based sensitivity analysis of the phonon Boltzmann transport equation (BTE) and a novel material interpolation technique, the ``transmission interpolation model'' (TIM). The key challenge in BTE optimization is handling the interplay between real- and momentum-resolved material properties. By parameterizing the material density with an \emph{interfacial} transmission coefficient, TIM is able to recover the hard-wall and no-interface limits, while guaranteeing a smooth transition between void and solid regions. We first use our approach to tailor the effective thermal-conductivity tensor of a periodic nanomaterial; then, we maximize classical phonon size effects under constrained diffusive transport, identifying a promising new thermoelectric material design. Our method enables the systematic optimization of materials for heat management and conversion and, more broadly, the design of devices where diffusive transport is not valid.}

%We introduce a methodology for density-based topology optimization of non-Fourier thermal transport in nanostructures, based upon the adjoint phonon Boltzmann transport equation (BTE) and a novel }. To this end, we also develop the transmission interpolation model (TIM), 




\maketitle
\section{Introduction}
Designing a nanomaterial with prescribed thermal properties is critical to many applications, such as heat management and thermoelectrics~\cite{Vineis,cahill2003nanoscale}. However, heat-conduction optimization in nanostructures remains challenging: Fourier's law breaks down~\cite{chen2021non}, heat transport becomes nonlocal, and standard topology-optimization methods~\cite{sigmund2013topology} for diffusive theories~\cite{dede2009multiphysics,haertel2015topology} are not readily applicable. An early study~\cite{evgrafov2009topology} developed the adjoint phonon Boltzmann transport equation (BTE) to design a material with a prescribed difference of temperature between two given points; in Ref.~\cite{evgrafov2009topology}, the local material density was related to the bulk phonon mean-free-path (MFP), a method that was proven successful for boundary conditions applied to influx phonon flux. However, such an approach is not suitable when shape optimization includes arbitrary adiabatic boundaries, a scenario that presents a challenge on its own: How to \emph{interpolate} a material so that phonons are scattered back isotropically at adiabatic walls (assuming diffuse boundaries), while also recovering the no-interface limit for uniform densities? We tackle this challenge by introducing the ``transmission interpolation model'' (TIM). The key concept behind TIM is that instead of relating the local density to volume-based parameters, such as the MFP, TIM paremetrized the material density in terms of phonon interfacial transmission.
 
In our implementation, we combine a BTE solver (Sec.~\ref{sec:gray}) with TIM (Sec.~\ref{sec:interpolation}), and chained them into a reverse-mode automatic differentiation pipeline~\cite{jax2018github}, which also includes density filtering and projection~\cite{sigmund2013topology} (Sec.~\ref{sec:optimization}). We apply our methodology to obtain new solutions to two exemplary problems: designing an anisotropic thermal-conductivity tensor in a periodic nanomaterial (Sec.~\ref{case1}) and, for thermoelectric applications~\cite{Vineis}, minimizing thermal transport while \emph{simultaneously} maintaining high electrical conductivity (Sec.~\ref{case2}). (To the latter end, we assume charge transport to be diffusive and thus implement a differentiable Fourier solver.) Several technical aspects, including the matrix-free solution of the BTE solver and the relationship between its forward and adjoint counterparts, are reported in the Appendices. The code developed for this work will be released in the OpenBTE package~\cite{romano2021openbte}.

Nondiffusive thermal transport, investigated from both theoretical~\cite{Ziman2001} and experimental~\cite{Lee2015BallisticSilicon,hochbaum2008enhanced,song2004thermal} standpoints, has opened up exciting engineering opportunities; however, it has also made modeling heat transport computationally challenging. One key departure from familiar Fourier diffusion is that phonons must be tracked in \emph{momentum} as well as position space, dramatically increasing the number of unknowns~\cite{Ziman2001}. If forward modeling is challenging, inverse design is even more difficult. In addition to Ref.~\cite{evgrafov2009topology}, mentioned above, there have only been a few studies aiming at gradient-based optimization of nanoscale thermal transport. For example, in a recent preprint~\cite{chen2022panoramic}, the adjoint BTE was used in conjunction with experiments to estimate phonon-related material properties. However, none of these works focus on systems with arbitrary adiabatic boundaries.
%both within gradient-free approaches~\cite{wei2020genetic,xiao2021inverse}, and they have mostly been based on gradient-free approaches that cannot handle large parameter spaces~\cite{Sigmund2011}.
In the simpler diffusive regime, density-based topology optimization has been routinely applied to macroscopic heat-transport problems~\cite{gersborg2006topology,zhang2008design,haertel2015topology,imediegwu2022multiscale,song2006evaluation}. The basic idea of density-based topology optimization~\cite{sigmund2013topology} is that each point in space, or each ``pixel'' in a discretized solver, is linked to a fictitious density $\rho(\mathbf{r})$ which is continuously varied between 0~and~1, representing two physical materials at the extremes, to optimize some figure of merit such as thermal conductivity.  Filtering and projection regularization steps~\cite{sigmund2013topology} ensure that the structure eventually converges to a physical material everywhere in the design domain, and a variety of methods are available to impose manufacturing constraints such as minimum lengthscales~\cite{zhou2015minimum,lazarov2016length}. Adjoint-based sensitivity analysis allows such huge parameter spaces to be efficiently explored~\cite{Sigmund2011}, enabling the computational discovery of surprising non-intuitive geometries. For instance, in Ref.~\cite{gersborg2006topology}, a heat-conducting material was designed to generate the least amount of heat under volume constraints. In that work, which mirrored the search for minimum-compliance materials for mechanical problems~\cite{sigmund200199}, the material density at each pixel could be directly related to the local \textit{bulk} thermal conductivity. In contrast, such a local relationship does not hold for the BTE. However, the BTE supports the use of transmission coefficients associated with the interfaces between dissimilar materials~\cite{chenbook}. In our work, therefore, we turn these coefficients into intermediate variables linking the material density to the phonon distributions using our TIM approach.

\section{The 2D single-MFP BTE}\label{sec:gray}

We are interested in computing the effective thermal conductivity tensor $\kappa$ of a periodic nanostructure. To this end, we consider a simulation domain composed of a square with side $L$, to which periodic boundary conditions are applied along both axes (see Fig.~\ref{fig0}-a). To calculate, for example, $\kappa_{xx}$, we apply a temperature jump of $\Delta T_{\mathrm{ext}}$ = 1 K across the $x$-axis, and average the $x$-component of heat flux,
\begin{equation}\label{eq:kappa_general}
\kappa_{xx} = -\frac{1}{\Delta T_{\mathrm{ext}}}\int_{-L/2}^{L/2}\mathbf{J}(L/2,y)\cdot \mathbf{\hat{x}}dy.
\end{equation} 
%Ref.~\cite{jbeili2021generalized}.
To calculate the heat flux, we note that, at the nanoscales, heat conduction deviates from the standard Fourier law because the mean-free-path (MFP) of heat carriers, i.e. \emph{phonons}, becomes comparable with the material's feature size. This phenomenon, commonly known as classical phonon size effects~\cite{chenbook}, can be captured by the phonon Boltzmann transport equation (BTE)~\cite{chenbook,peierls1929kinetischen,romano2021efficient}.
%can be described by the phonon BTE, which is an integro-differential equation that takes into account 
%describes heat conduction beyond the diffusive regime~\cite{chenbook,peierls1929kinetischen}. By taking into account the phonons' mean-free-path (MFP) distribution, which is in general anisotropic, it captures the mode-resolved interaction between the geometry and the underlying material~\cite{romano2021efficient}, a phenomenon commonly known as classical phonon size effects~\cite{chenbook}.
There are different flavors of the BTE, depending on the needed accuracy. In this work, we use the single-MFP version of the BTE, a textbook-case also known as the \textit{gray} model~\cite{chenbook}; within this approximation, a bulk material is simply parameterized by its thermal conductivity $\kappa_{\mathrm{bulk}}$ and MFP $\Lambda$.  We consider two-dimensional (2D) transport, i.e. phonon directions are parameterized by the polar angle $\phi$. With these assumptions, the gray BTE reads as
\begin{eqnarray}\label{eq:bte_gray}
    \Lambda\mathbf{\hat{s}}(\phi)\cdot \nabla \g(x,y,\phi) +  \g(x,y,\phi) =\frac{1}{2\pi}\int_{-\pi}^{\pi} \g(x,y,\phi' )d\phi',
\end{eqnarray}
where $\g(x,y,\phi) = \left[T(x,y,\phi)-T_0\right]/\Delta T_{\mathrm{ext}}$ is a deviational pseudo phonon temperature, normalized by $\Delta T_{\mathrm{ext}}$ (in short, ``phonon temperatures'' throughout the text), the unknown of our problem; $T_0$ is a reference temperature. The vector $\mathbf{\hat{s}}(\phi)=\sin{\phi}\mathbf{\hat{x}}+\cos{\phi}\mathbf{\hat{y}}$ is the phonon direction, illustrated in Fig.~\ref{fig0}a. Note that the BTE is often formulated in terms of distribution functions or energy density~\cite{chenbook,Majumdar1993am,murthy1998finite}. The temperature formulation used here is simply obtained by a change of variables~\cite{romano2015}. Lastly, the angular-resolved heat flux is given by
\begin{equation}\label{eq:thermal_flux}
    \mathbf{J}(x,y,\phi)=\frac{2\kappa_{\mathrm{bulk}}}{\Lambda} \g(x,y,\phi)\hat{\mathbf{s}}(\phi),
\end{equation}
 with the total heat flux being $\mathbf{J}(x,y) = \left(2\pi\right)^{-1}\int_{-\pi}^{\pi} \mathbf{J}(x,y,\phi) d\phi$.
Although here we employ a simplified version of the BTE, the developed methodology can be readily applied to more sophisticated versions. Combining Eqs.~\eqref{eq:kappa_general}and~\eqref{eq:thermal_flux}, we define the \emph{normalized} effective thermal conductivity tensor, $\bar{\kappa}_{xx} = \kappa_{xx}/\kappa_{\mathrm{bulk}}$, as
\begin{equation}\label{eq:kappa_bte}
    \bar{\kappa}_{xx} = -\frac{1}{\Lambda \pi}\int_{-\pi}^{\pi} \int_{-L/2}^{L/2}  \g(L/2,y,\phi) \mathbf{\hat{s}}(\phi)\cdot \mathbf{\hat{x}}dy d\phi.
\end{equation}
Similarly, $\bar{\kappa}_{yy}$ is evaluated by applying a temperature gradient along the $y$-axis. Throughout this work we use $\Lambda = L$, thus neither of these two values need to be specified in absolute values. (Note that this simplification does not hold for nongray materials, where $L$ needs to be specified in physical units.) Analogously, thanks to linearity, we don't need to provide explicit values for $\kappa_{\mathrm{bulk}}$ and $T_0$. Internal boundaries of the simulation domain are modeled as diffuse hard-walls, i.e. phonons approaching the surface are scattering back isotropically~\cite{Ziman2001,murthy2002numerical}. %Combining the energy conservation requirement, i.e. $\mathbf{J}\cdot \mathbf{\hat{n}} = 0$ ($\mathbf{\hat{n}}$ is the normal of the surface), and Eq.~\ref{eq:thermal_flux}, a diffuse surface translates into imposing the following \emph{boundary} temperature to phonons leaving the surface: 
%\begin{equation}\label{eq:TB}
%\g^B(\mathbf{r}^-) = \frac{1}{2}\int_{\s \cdot \n > 0}\g (\mathbf{r}^-,\phi)\mathbf{\hat{s}}(\phi)\cdot \mathbf{\hat{n}}d\phi,
%\end{equation}
%where we used $\int_{\s \cdot \n > 0}\mathbf{\hat{s}}(\phi)\cdot \mathbf{\hat{n}}d\phi = 2$. 
In Sec.~\ref{sec:interpolation}, we will describe this boundary condition as the hard-wall limit of interpolation method used to account for phonon transport in arbitrary material distribution. Equation~\eqref{eq:bte_gray} is discretized using the finite-volume approach both in real- and angular-space. The resulting linear system reads 
\begin{equation}\label{eq:bte_discretized}
    \sum_{\mu' c'} G_{\mu c}^{\mu' c'} \g_{\mu' c'} = P_{\mu c},
\end{equation}
where $\mu$ and $c$ label angular and real-space, respectively. Equation~\eqref{eq:bte_discretized} is solved using a matrix-free Krylov subspace method. The expressions for the terms $G_{\mu c}^{\mu'c'}$ and $P_{\mu c}$, as well as details on the iterative solution of Eq.~\eqref{eq:bte_discretized}, are provided in Sec.~\ref{sec:bte_solver}.

%are provided in Sec.~\ref{sec:bte_solver}. As detailed in Sec.~\ref{sec:bte_solver}, Eq.~\eqref{eq:bte_discretized} is first \emph{flattened} and then solved using a matrix-free Krylov subspace method.
%For $\Lambda\rightarrow 0$, Eq.~\eqref{eq:bte_gray} recovers Fourier's law, i.e. $\nabla \cdot \left[\kappa(x,y)\nabla \g(x,y)\right] = 0$, for which the effective thermal conductivity tensor is computed by
%\begin{equation}
%    \bar{k}^F_{xx} = -\int_{-L/2}^{L/2} \kappa(L/2,y) \nabla \g(x,y)\cdot \mathbf{\hat{x}} dy
%\end{equation}
Lastly, we note that in this work a Fourier solver is also used, where the temperature is only described in real-space. We will refer to the corresponding normalized effective thermal conductivity as $\bar{\kappa}^{\mathrm{F}}$. In this case, the linear system to solve is $\sum_{c'}A_{cc'}\g_{c'}=b_{c}$. The expressions for $A_{cc'}$ and $b_c$, as all as details on gradient calculations of the Fourier solver are reported in Sec.~\ref{sec:fourier_solver}.

 %in our case,  in our case between a hard wall and no interface at all.
\section{The Transmission Interpolation Model}\label{sec:interpolation}

Density-based topology optimization requires a differentiable transition between material properties~\cite{sigmund2013topology}; that is, one must be able to deal with arbitrary material distributions described by a fictitious density $\boldsymbol \rho\in [0,1]^N$, where $N$ is the number of ``pixels'' (design degrees of freedom) in the material~\cite{bendsoe1999material}. Following Fig.~\ref{fig0}-b, we begin by considering an interface between two pixels, with different densities, $\rho_1$ and $\rho_2$. The interface between them has normal $\n$ pointing toward pixel 2. Furthermore, we define the phonon temperatures in those two pixels as $\g_1(\phi)$ and $\g_2(\phi)$. Note that while we have discretized the real space, in this section we use a continuous representation for $\phi$. In our case, a material interpolation model must satisfy two limit cases: When $\rho_1 = \rho_2$, there should be no extra phonon scattering across their interface; on the other side, when $\rho_1 = 0$ and $\rho_2 = 1$, phonons must scattered back isotropically toward region 1.

A possible material interpolation model is given in Ref.~\cite{evgrafov2009topology}, where the MFP depends on the material density through $\Lambda^{-1}(\rho) = \Lambda_a^{-1}\rho + (1-\rho) \Lambda_b^{-1}$, with $\Lambda_a$ and $\Lambda_b$ associated to two different phases. This approach was successfully applied for boundary conditions on incoming phonon flux. However, it may be problematic for the adiabatic hard-wall limit, as explained in the following. Adopting the approach from Ref.~\cite{evgrafov2009topology}, the heat flux at the interface between the two pixels is
\begin{eqnarray}\label{eq:condition}
    \mathbf{J}(\mathbf{r})\cdot\n &=&\frac{2\kappa}{\pi} \left[\frac{\rho_1}{\Lambda_a} + \frac{(1-\rho_1)}{\Lambda_b} \right]\int_{\s\cdot\n\ge0}\g_1(\phi) \s\cdot\mathbf{\hat{n}}d\phi+\nonumber \\ &+&\frac{2\kappa}{\pi}\left[\frac{\rho_2}{\Lambda_a} + \frac{(1-\rho_2)}{\Lambda_b} \right]\int_{\s\cdot\n<0}\g_2(\phi) \s\cdot\n d\phi.
\end{eqnarray}
 For adiabatic boundaries, we may assign $\Lambda_a$ to the solid phase and a fictitious $\Lambda_b = \infty$ to the void one; in this case, the second part of Eq.~\eqref{eq:condition} will be zero because $\rho_2$ = 0 but the first part (which has $\rho_1 = 1$) will be different than zero. Consequently, such an approach would lead to a nonzero net thermal current, while we wish to have an adiabatic surface. Note that this conclusion applies to generic adiabatic surfaces within the context of the BTE and is not tied to our choice of diffuse scattering. 

To lift these limitations, we attack the problem from a different angle: We parameterize the material density via a phonon transmission coefficient~$t$. In doing so, we borrow a methodology developed for thermal transport across dissimilar materials, where the transmission coefficient is used to impose the distributions \textit{leaving} the interface~\cite{singh2011effect}. Specifically, we introduce the boundary conditions 
\begin{eqnarray}\label{eq:bc}
\g_1(\phi)=t\g_2(\phi) + (1-t)\g^B_{12},  \,\,\,\,\, \mathrm{for} \,\,\phi: \s\cdot \n < 0 \nonumber \\
\g_2(\phi)=t\g_1(\phi) + (1-t)\g^B_{21},\,\,\,\,\, \mathrm{for} \,\,\phi: \s\cdot \n \ge 0,
\end{eqnarray}
where $\g^B_{ij}$ is the boundary temperature at the interface between pixels i and j, thermalizing phonons traveling into pixel i. Its expression is given by
\begin{eqnarray}\label{eq:TB}
    \g^B_{12} &=& \frac{1}{2}\int_{\s \cdot \n \ge 0}\g_1 (\phi)\mathbf{\hat{s}}(\phi)\cdot \mathbf{\hat{n}}d\phi \nonumber \\
    \g^B_{21} &=& - \frac{1}{2}\int_{\s \cdot \n < 0}\g_2 (\phi)\mathbf{\hat{s}}(\phi)\cdot \mathbf{\hat{n}}d\phi,
\end{eqnarray}
where we used $\int_{\s \cdot \n \ge 0}\mathbf{\hat{s}}(\phi)\cdot \mathbf{\hat{n}} d\phi = -\int_{\s \cdot \n < 0}\mathbf{\hat{s}}(\phi)\cdot \mathbf{\hat{n}} d\phi = 2$. The term $t$ in Eq.~\ref{eq:bc} is a transmission coefficient, which we define as
\begin{equation}\label{eq:transmission}
    t = 2\frac{\rho_1\rho_2}{\rho_1+\rho_2}.
\end{equation}
\begin{figure}[ht]
\begin{center}
{\includegraphics[width=0.75\textwidth]{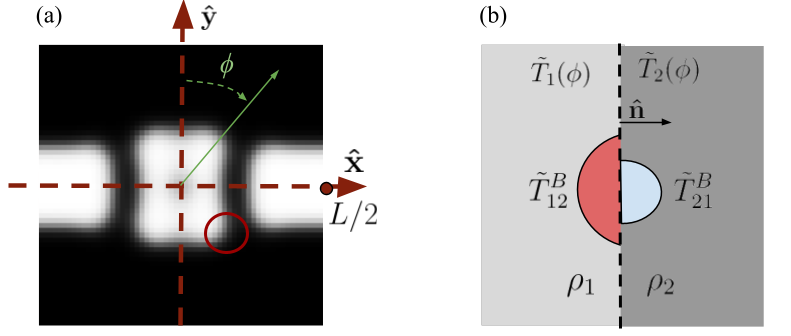}}
\caption{a) The simulation domain: a square centered in (0,0) and with side $L$. b) An example of interface between two regions with different densities, where the impinging, transmitted and isotropically reflected phonon temperatures are also shown.}\label{fig0}
\end{center}
\end{figure}
It is straightforward to show that if either $\rho_1$ or $\rho_2$ is zero, then the RHS of Eq.~\ref{eq:bc} reduces to the hard-wall case. On the other side, if $\rho_1 = \rho_2$, there will be no interface. To summarize, our parametrization does not relate the material density to a \emph{bulk-like} property (such as the MFP from Ref.~\cite{evgrafov2009topology}), but rather to the amount of incoming flux. To distinguish this approach from traditional material interpolation methods, we name it the ``Transmission Interpolation Model'' (TIM). In passing, we note that transmission coefficients of the form $t^\gamma$, with $\gamma > 1$ would also be a suitable interpolation approach. However, investigating this more general case is outside the scope of our work. Details on the angular discretization of TIM is reported in~Sec. \ref{sec:bte_solver}. 

\section{The optimization pipeline}\label{sec:optimization}

In this section, we outline the method for computing $\nabla_{\boldsymbol \rho} \bar{\kappa}$, which will be used in our optimization algorithm. We begin by noting that density-based topology optimization presents two major challenges: The emergence of rapidly oscillating ``checkerboard'' patterns that fail to converge with increasing spatial resolution; and gray ($0 < \rho < 1$) pixels, to which no physical material can be associated~\cite{bendsoe2003topology}.
\begin{figure*}[t!]
\begin{center}
{\includegraphics[width=0.95\textwidth]{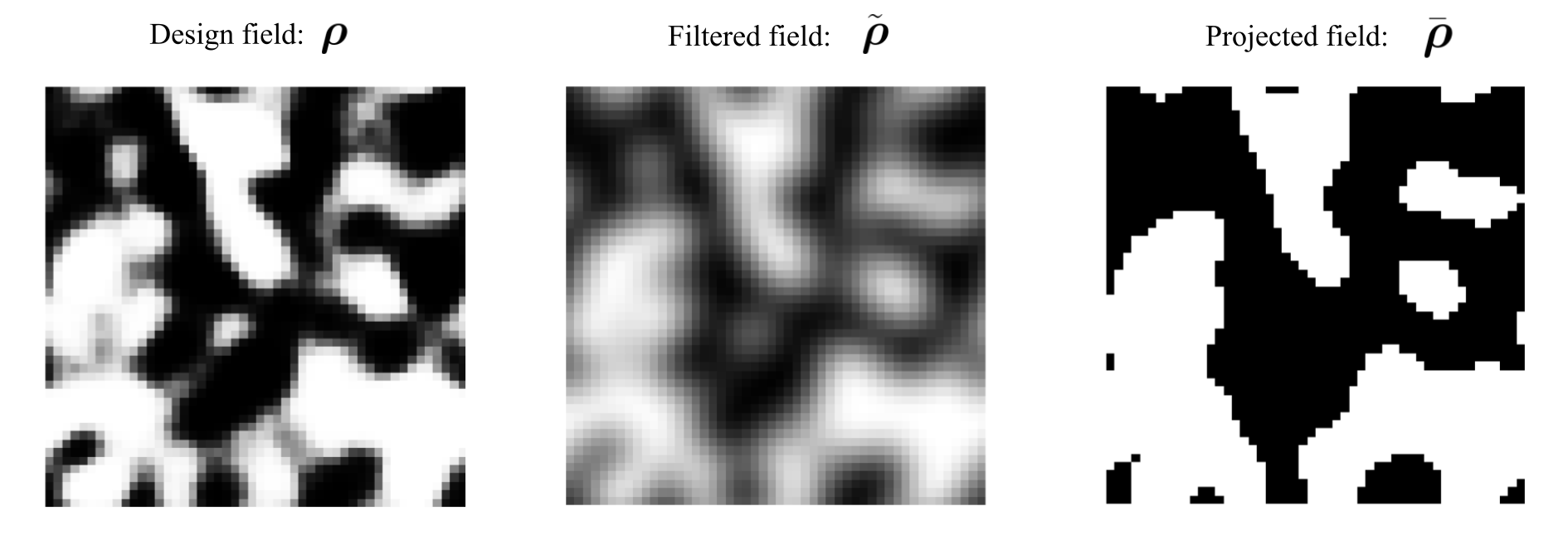}}
\caption[]{The design field $\boldsymbol \rho$ (in the left panel) is the optimization parameter. The filtered field  $\tilde{\boldsymbol \rho}$, in the middle panel, is computed after the convolution with a conic filter, with Kernel defined in Eq.~\ref{kernel}. The projected filter $\bar{\boldsymbol \rho}$, in the right panel, obtained with Eq.~\ref{tanh}, is the input to the BTE.}\label{fig:fig_new}
\end{center}
\end{figure*}
These two issues are commonly resolved using filtering and thresholding, respectively~\cite{sigmund2013topology}. As shown in Fig.~\ref{fig:fig_new}, given a design density $\boldsymbol \rho$ (for convenience, from now on, we will work with a discretized domain), we first filter it,  $\tilde{\boldsymbol \rho} = \mathbf{w}* \boldsymbol \rho $, where, in this case, $\mathbf{w}$ is a conic filter with radius $R$, 
\begin{equation}\label{kernel}
w_c =  \begin{cases}
 \frac{1}{a}\left(1-\frac{\mid \mathbf{r}_c\mid}{R}\right),& \mid \mathbf{r} \mid <R\\
    0,              & \text{otherwise}.
\end{cases}
\end{equation}
In Eq.~\eqref{kernel}, $a$ is a normalization factor ($= \pi R^2/3$ in the continuum limit), $\mathbf{r}_c$ is the centroid of the grid point $c$, and $R$ is the radius of our filter. In this work, $R = L/10$. The thresholding, $\bar{\boldsymbol \rho} = f_p(\tilde{\boldsymbol \rho})$, is then carried out using the following function~\cite{wang2011projection}
\begin{equation}\label{tanh}
    \bar{\boldsymbol{\rho}}=\frac{\tanh{\beta \eta}+\tanh(\beta(\tilde{\boldsymbol{\rho}}-\eta))}{\tanh{\beta \eta}+\tanh(\beta(1-\eta))},
\end{equation}
where $\eta$ and $\beta$ are threshold parameters. The resulting field, referred here as ``projected'' is, therefore, used directly by the BTE solver; in this work, we use $\eta=0.5$; the term $\beta$, on the other side, is increased during the optimization procedure~\cite{hammond2021photonic}, in order to guarantee a good degree of topology variability (especially early on in the optimization process) while ensuring a final binary structure. In this work, we start with $\beta = 2$ and double it every 20 iterations, until convergence is reached.

Once the relationship $\bar{\boldsymbol {\rho}}(\boldsymbol \rho)$ is implemented, we can use the chain rule
\begin{equation}\label{eq:kappa_bte_grad}
   \frac{d \bar{\kappa}}{d \rho_p} = \sum_{\mu cp'}\frac{\partial \bar{\kappa}}{\partial \g_{\mu c}}\frac{\partial \g_{\mu c}}{\partial   \bar{\rho}_{p'}}\frac{\partial \bar{\rho}_{p'} }{\partial \rho_p }+\sum_{p'} \frac{\partial \bar{\kappa}}{\partial \bar{\rho}_{p'}}\frac{\partial \bar{\rho}_{p'}}{\partial \rho_p},
\end{equation}
which is evaluated using reverse-mode automatic differentiation, implemented in JAX~\cite{jax2018github}. Specifically, for $\partial \bar{\kappa}/\partial \bar{\rho}_p$ we use the \emph{adjoint} method~\cite{strang2007computational}, which allows to compute such a gradient by solving the linear system
\begin{equation}\label{eq:adjoint_ori}
   \sum_{\mu'c'}  G_{\mu' c' }^{\mu c}\g_{\mu'c'}^{\mathrm{adj}} =  P^{\mathrm{adj}}_{\mu c},
\end{equation}
with $P^{\mathrm{adj}}_{\mu c}$ defined in Sec.~\ref{sec:bte_solver}. In practice, we use the relationship
\begin{equation}\label{eq:bte_symmetry1}
\g^{\mathrm{adj}}_{\mu c} = - \g_{-\mu c},
\end{equation}
derived in Sec.~\ref{sec:bte_solver}. Therefore, the adjoint solution is computed by post-processing the solution of Eq.~\eqref{eq:bte_discretized}, achieving a significant boost in computational efficiency. 

Furthermore, as $\nabla_{\bar{\boldsymbol \rho}}\bar{\kappa}$ is available at each iteration while solving the forward problem, we adopt an early termination criteria, based on $\bar{\kappa}$ \emph{and} $\mid\nabla_{\bar{\boldsymbol \rho}}\bar{\kappa}\mid$. This approach extends Ref.~\cite{amir2010efficient}, where early termination strategies were based on the error on the objective function alone.
%iterative approaches, when used in the context of topology optimization, may benefit from  In this work, we extend this approach by also considering the error on $\mid\nabla_{\bar{\boldsymbol \rho}}\bar{\kappa}\mid$; this method is facilitated by the fact that at each iteration of the linear solver, we have $\g_c$ \emph{and} $\g^{\mathrm{adj}}_c$, through Eq.~\eqref{eq:bte_symmetry1}. 
The sensitivity of $\bar{\kappa}$ with respect to the projected density is provided through the custom vector-Jacobian-product $\mathrm{vJp}(a) = a \partial \bar{\kappa}/\partial \rho_p$. Lastly, we note that for the Fourier solver, the forward and adjoint solutions are related $\g_c^{\mathrm{adj}} = - \g_c$, as derived in Sec.~\ref{sec:fourier_solver} Similarly to the BTE case, we use this relationship to avoid solving the adjoint problem.

\section{Case I: Tailoring the Effective Thermal Conductivity Tensor}\label{case1}

In this section, we show an example of how topology optimization may be employed to design a periodic material with a prescribed effective thermal conductivity tensor, $\tilde{\kappa}$, and with a porosity larger than $\tilde{\varphi}$. To this end, we define the objective function
\begin{equation}
    g(\boldsymbol\rho) =  
\mid \bar{\kappa}(\boldsymbol\rho) - \tilde{\kappa}\mid_{\mathrm{Fro}} = \sqrt{\Delta\kappa_{xx}(\boldsymbol\rho)^2 +\Delta\kappa_{yy}(\boldsymbol\rho)^2 },
\end{equation}
where $||.||_{\mathrm{Fro}}$ is the Frobenius norm, and $\Delta \kappa_{ii}(\boldsymbol \rho) = \bar{\kappa}_{ii}(\boldsymbol \rho)-\tilde{\kappa}_{ii}$. The effective thermal conductivity tensor,
\begin{equation}
    \bar{\kappa}(\boldsymbol \rho) = \begin{pmatrix}
\bar{\kappa}_{xx}(\boldsymbol\rho) & 0 \\
0 & \bar{\kappa}_{yy}(\boldsymbol\rho)
\end{pmatrix},
\end{equation}
is evaluated by solving Eq.~\eqref{eq:bte_gray}, for each perturbation direction, \textit{after} filtering and projecting. The sensitivity of the objective function is
\begin{equation}
\nabla_{\boldsymbol\rho}g(\boldsymbol\rho) = g(\boldsymbol\rho)^{-1}\left[ \Delta\kappa_{xx}(\boldsymbol\rho)\nabla_{\bar{\boldsymbol\rho}}\bar{\kappa}_{xx}(\boldsymbol\rho) +\Delta\kappa_{yy}(\boldsymbol\rho)\nabla_{\boldsymbol\rho}\bar{\kappa}_{yy}(\boldsymbol\rho)\right],
\end{equation}
where the terms $\nabla_{\boldsymbol\rho}\bar{\kappa}_{ii}(\boldsymbol\rho)$ are computed using Eq.~\ref{eq:kappa_bte_grad}. The above discussion allows us to lay out the optimization algorithm 
\begin{eqnarray}\label{opt1}
    \min_{\boldsymbol\rho}g(\boldsymbol\rho)\nonumber\\
    0 \le \boldsymbol\rho \le 1 \nonumber\\
    \mathrm{s.t.} \sum_n \bar{\rho}_n \le (1- \tilde{\varphi}) N,
\end{eqnarray}
where $N$ is the number of pixels. In this section, the chosen porosity is $\tilde{\varphi} = 0.25$. As the optimizer, we use an open-source implementation~\cite{johnson2014nlopt} of the method of moving asymptotes (MMA)~\cite{svanberg2002class}, which converges globally (i.e. it guarantees to find a local minimum from every starting point). 

\begin{figure*}[t!]
\begin{center}
{\includegraphics[width=0.75\textwidth]{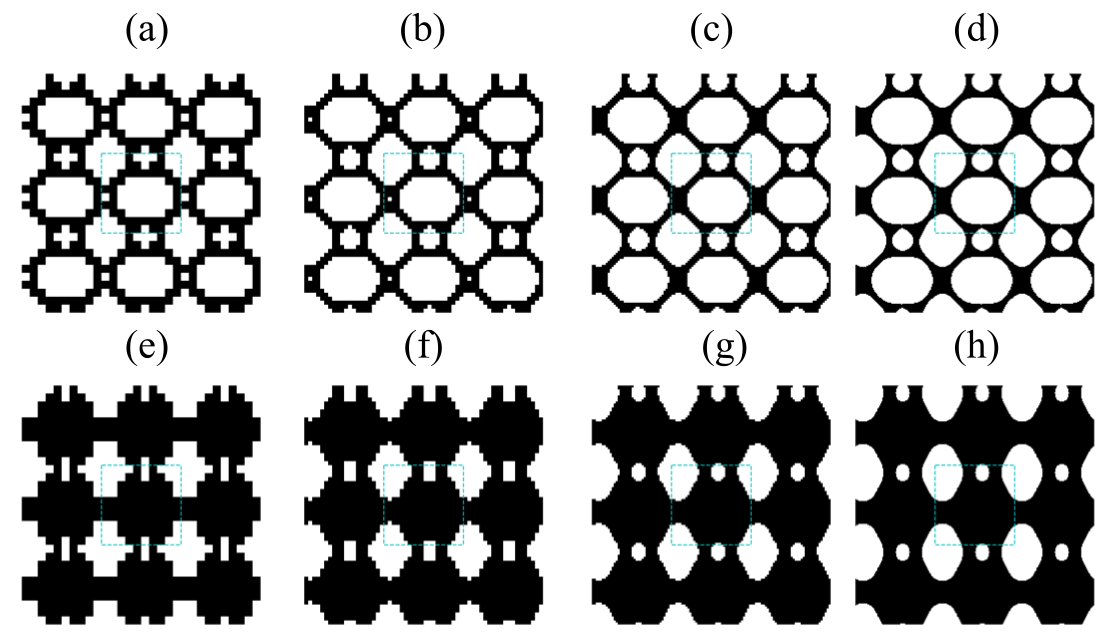}}
\caption{The optimized structures using the Fourier solver, for grid size (a) $20\times 20$, (b) $40\times 40$, (c) $60\times 60$ and (d) $80\times 80$. Similarly, the final structures for the BTE case are illustrated in (e), (f), (g) and (h). The dotted square represent the unit cell, whose width is $L=\Lambda$. The filter's radius is $R=L/10$.}\label{fig:fig00}
\end{center}
\end{figure*}

As a first example, we choose $\tilde{\kappa}_{xx}=\tilde{\kappa}_{yy} = 0.15$. To ensure mesh convergence on a particular local minimum, we use the following algorithm: 
\begin{itemize}
    \item[] 1. Optimize a structure at a coarse resolution using a random configuration as the initial structure.
    \item[] 2. Upsample the optimal structure by doubling the resolution.
    \item[] 3. Optimize a structure using the configuration created in step 2 as the first guess. Note that the filter's radius does not change in \textit{physical} units, but doubles in \textit{pixel} units.
    \item[] 4. Repeat from step 2.
\end{itemize}
Figure~\ref{fig:fig00} illustrates the optimized structures for grid sizes $N=10\times 10, 20\times 20,40\times 40$ and $80\times 80$. For both the Fourier and BTE cases, a shape convergence is achieved.  For the rest of this study we adopt a grid of $60\times 60$, using a random configuration as a first guess. A striking differences between the two solvers is that for the BTE case the pattern is coarser. In fact, phonon size effects are known to be more effective than macroscopic reduction with the same geometric constraints~\cite{sharvin1965possible}.

We now turn to the design an anisotropic material. Thermal anisotropy may be induced by boundary engineering, even though the base material is isotropic. Symmetry-breaking boundaries are effective at all scales, although it has been shown numerically that nanostructuring may enhance anisotropy with respect to the macroscopic counterpart~\cite{romano2017thermal}. In this example, we choose $\bar{\kappa}_{xx} = 0.3$ and $\bar{\kappa}_{yy} = 0.1$, with the resulting anisotropy $ \bar{\kappa}_{xx}/\bar{\kappa}_{yy}= 3$.

\begin{figure*}[t!]
{\includegraphics[width=0.95\textwidth]{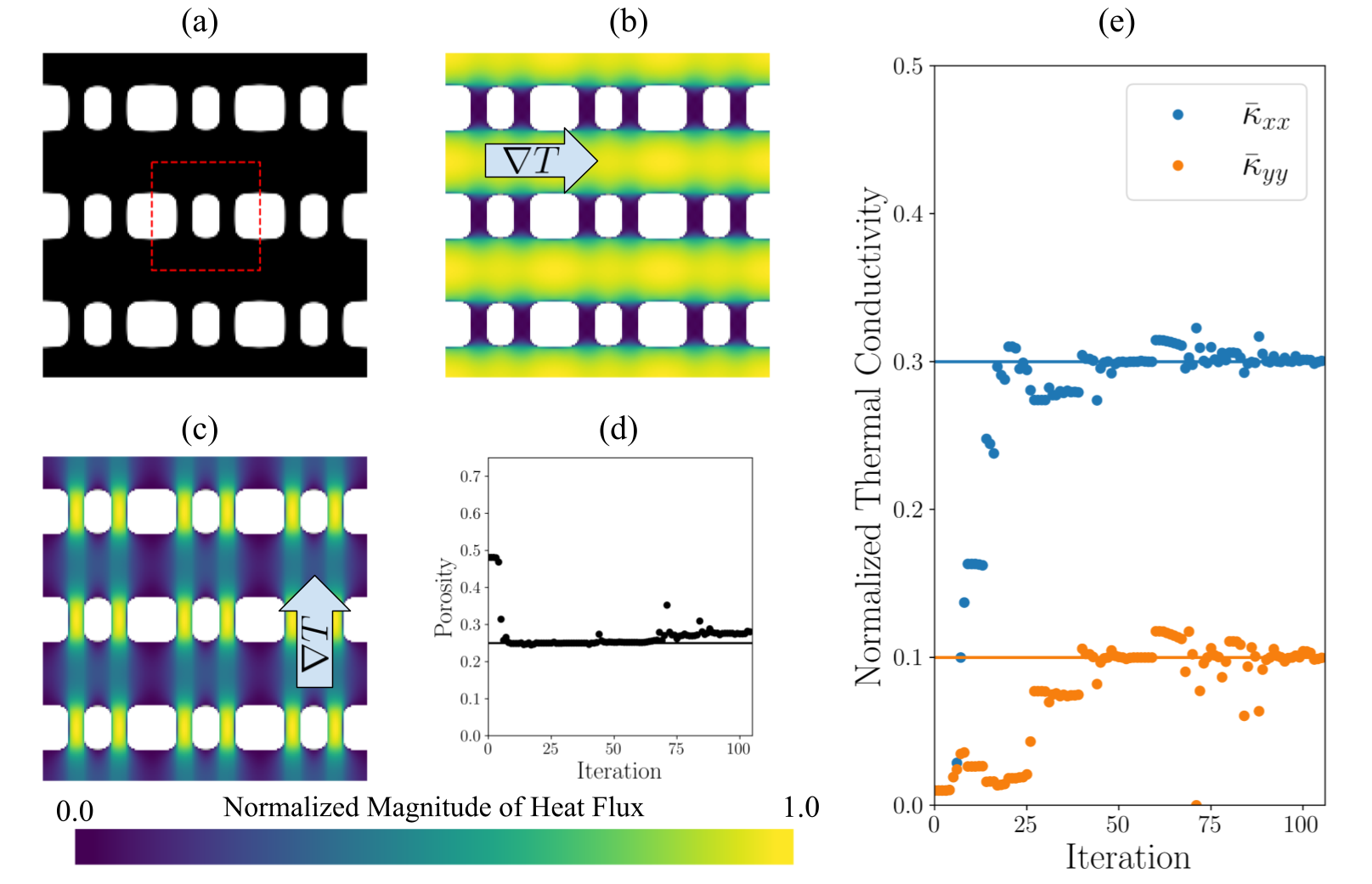}}
\caption{a) The optimized structure after solving Eq.~\eqref{opt1}, e) The evolution of the effective thermal conductivity tensor. The dotted lines are the desired value of the two components, d) the evolution of the porosity. The dotted line is the lower bound imposed by the inequality constraint. Normalized magnitude of thermal flux when the gradient is applied along the $x$-axis (b) and $y$-axis (c).}\label{fig2}
\end{figure*}
Convergence is reached within 100 iterations. The final structure, shown in Fig.~\ref{fig2}-a, is made by two types of pores, which block heat along \textit{y} more effectively than along \textit{x}. This effect is exemplified by the magnitude of thermal flux shown in Fig.~\ref{fig2}-b and Fig.~\ref{fig2}-c, for $\bar{\kappa}_{xx}$ and $\bar{\kappa}_{yy}$, respectively. Note that the final values obtained with Fourier's law are $\bar{\kappa}^F_{xx}\approx$ 0.61 and $\bar{\kappa}^F_{yy}\approx$ 0.46, with anisotropy 1.33, well below the prescribed value.

\section{Case II: Maximixing Phonon Scattering}\label{case2}

\begin{figure*}[t!]
{\includegraphics[width=0.95\textwidth]{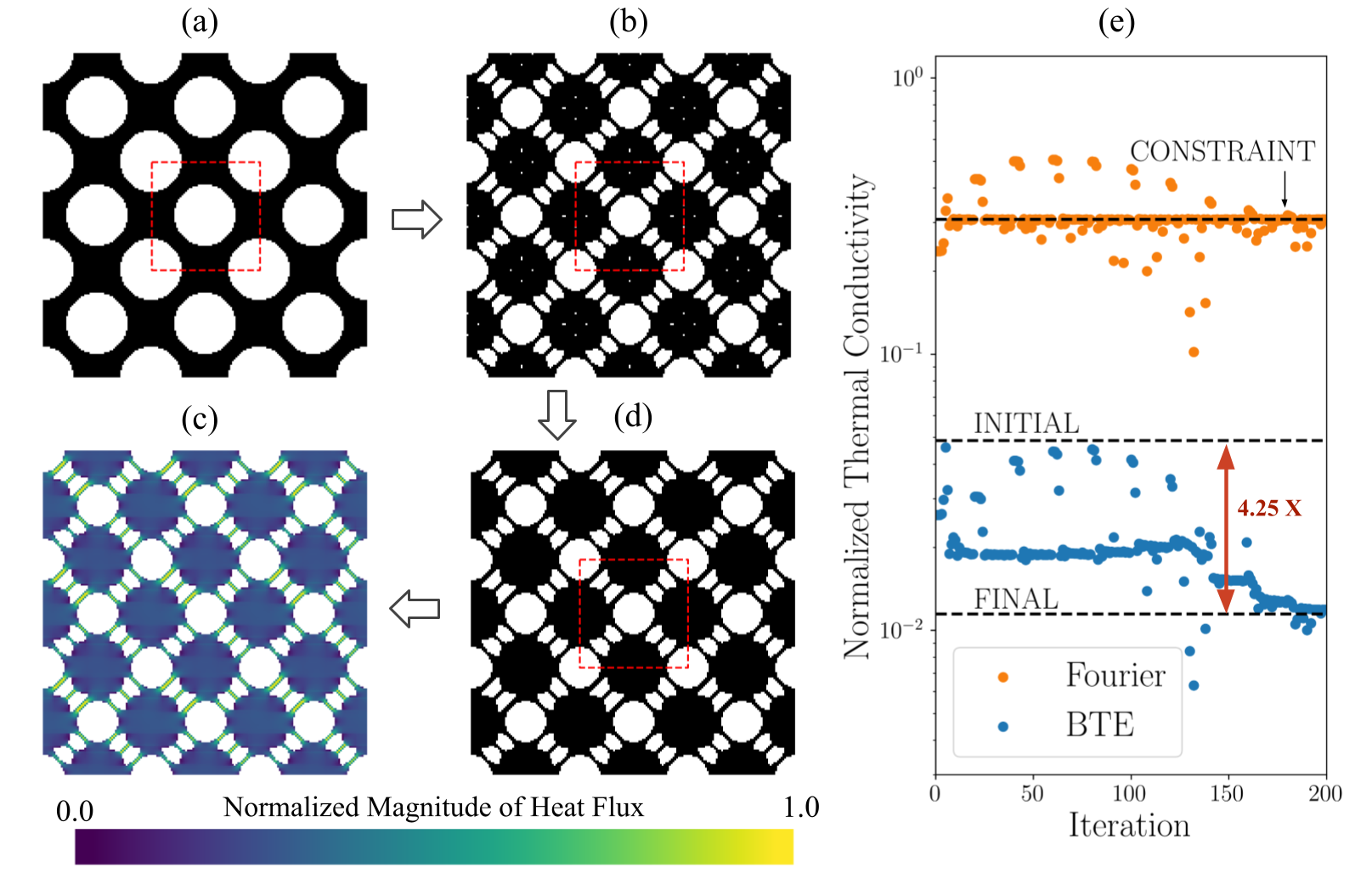}}
\caption{The a) initial, b) optimized and d) polished structure in relation to the optimization problem described by Eq.~\ref{opt2}. e) The evolution of $\bar{\kappa}$ and $\bar{\kappa}^{\mathrm{F}}$. c) The normalized magnitude of flux when a temperature gradient is applied along the $x$-axis. }\label{fig3}
\end{figure*}

In thermoelectric applications it is desirable to minimize thermal transport while not degrading the electrical conductivity~\cite{rowe2018crc} ($\sigma$). In fact, the thermoelectric figure-of-merit is given by $ZT=T S^2\sigma/\kappa$, where $S$ is the Seebeck coefficient. In highly-doped semiconducting nanostructures, these conditions can be met simultaneously due to the short phonon MFP compared to that of the electrons~\cite{Vineis,qiu2015first}. If the MFPs of the electrons are much shorter than the material's characteristic length, we may assume diffusive electronic transport. Consequently, minimizing the thermal conductivity while maintaining high diffusive transport is beneficial to ZT. Furthermore, in order to understand phonon scattering, most studies focus on the value of the effective thermal conductivity compared to that obtained with Fourier's law~\cite{Tang2010,lee2017investigation}. In passing, we note that macroscopic geometrical effects, often referred to as ``porosity factor''~\cite{verdier2017thermal}, in some cases have analytical solutions. For example, in aligned porous systems with circular pores and porosity $\varphi$, it has the analytical solution $\bar{\kappa}^{\mathrm{F}}=(1-\varphi)/(1+\varphi)$~\cite{hasselman1987effective}. We choose as baseline a porous material with staggered pores of circular shape~\cite{Romano2014,song2004thermal,anufriev2020ray}, as shown in Fig.~\ref{fig3}-a.
%This configuration has a smaller $\bar{\kappa}$ than that of systems with a square lattice of pores (provided the porosity is the same) because of the vanishing ``view factor''~\cite{Romano2014,song2004thermal,anufriev2020ray}. For this reason, such a configuration is often preferred. 
The chosen porosity is $\varphi=0.5$, to which it corresponds the isotropic tensors $\bar{\kappa}^{F}\approx 0.31$ and  $\bar{\kappa}\approx 0.049$. The goal of our optimization is, therefore, to achieve $\bar{\kappa} < 0.049$ under the constraint $\bar{\kappa}^{F} \ge 0.31$;  furthermore, we require $\bar{\kappa}$ to be isotropic. We use this baseline configuration as a first guess for our optimization algorithm, solving the problem:
\begin{eqnarray}\label{opt2}
\min_{\boldsymbol \rho} g(\boldsymbol \rho)\\
0 \le \boldsymbol\rho \le 1 \nonumber\\
\mathrm{s.t.}\,\,\, \bar{\kappa}^F_{xx}(\boldsymbol \rho) \ge 0.31\\
\mathrm{s.t.}\,\,\, \bar{\kappa}^F_{yy}(\boldsymbol \rho) \ge 0.31,
\end{eqnarray}
where
\begin{equation}\label{costfunction}
    g(\boldsymbol \rho) = \frac{1}{\sqrt{2}} \sqrt{\left(\bar{\kappa}_{xx}-\bar{\kappa}_{yy}\right)^2 + \bar{\kappa}_{xx}^2 + \bar{\kappa}_{yy}^2  }=\sqrt{\bar{\kappa}_{xx}^2+\bar{\kappa}_{yy}^2-\bar{\kappa}_{xx}\bar{\kappa}_{yy}},
\end{equation}
is the cost function to be minimized. We run this optimization problem with $\gamma = 3$ (see Sec.~\ref{sec:fourier_solver}), (smaller values would mostly lead to stagnation). Convergence is reached in 200 iterations, as shown in Fig.~\ref{fig3}-e. Remarkably, the optimized structure, shown in Fig.~\ref{fig3}-b, has an isotropic tensor of $\bar{\kappa}\approx 0.011$, roughly 4.25 times smaller than that of the baseline; yet, $\bar{\kappa}^{F}$ is right above the imposed constrain. The final porosity is 0.63. We point out the presence of small pores that are one or two pixels in size; to realize a structure that is more amenable from a manufacturing standpoint, we fill these small regions with solid phase, while making sure that the performance is not degraded (both $\bar{\kappa}$ and $\bar{\kappa}^{F}$ are within 1\% of those of the unpolished structure). The \textit{polished} configuration is shown in Fig.~\ref{fig3}-d. In passing, we note that it is possible to impose minimum-linewidth and minimum-linespacing conditions by adding differentiable inequality constraints to the optimization algorithm~\cite{hammond2021photonic,zhou2015minimum}, and in the future we plan to optimize the design for specific manufacturing processes in this way. Lastly, we recall that diffusive transport is scale-free, thus in principle we can begin from the staggered configuration and scale it down until we reach the same $\bar{\kappa}$ obtained from the optimization. However, such a configuration would be much more challenging to be manufactured (it would have several smaller pores) than the one depicted in Fig.~\ref{fig3}.

The optimized structure can be analyzed either from the void or the solid regions' point of view. In the former case, we have staggered pores with smaller void regions in between. More interestingly, from the solid regions' perspective we note a regular pattern of islands interconnected via three thin bridges on four opposite sides. As shown in Fig.~\ref{fig3}-c and as a consequence of energy conservation, heat flux peaks over these connections. The emergence of such a topology can be analyzed in terms of transport across a single orifice of width $a$. This problem was first investigated by Maxwell~\cite{maxwell1873treatise} in the diffusive regime, showing that the thermal resistance is proportional to $1/a$; on the other side, Sharvin~\cite{sharvin1965possible} predicted that in the ballistic regime, i.e.~for $\Lambda << a$ (such as in our case), the resistance goes as $\Lambda/a$. Thus, it is clear that thin channels are a promising platform for decoupling diffusive and nondiffusive transport. Both the abovementioned approximations assume infinite leads. A very recent study~\cite{spence2022phonon}, however, investigates heat transport across a single Si-based orifice using the BTE within a Monte-Carlo framework, revealing a significant role of the geometry of the orifice and leads (i.e.~the structures attached to the two ends of the channel) on the overall thermal resistance. Our optimization approach, therefore, automatically identifies a structure featuring orifices, while concurrently optimizing the  geometries of the leads.
            
\section{Conclusions}

In this work, we develop a model, termed the Transmission Interpolation Model (TIM), that is able to smoothly interpolate material properties in the context of nondiffusive heat transport. The key concept behind TIM is that instead of linking a volume-based quantity to the material density, e.g. the bulk thermal conductivity, it parametrizes an \emph{interfacial} transmission coefficient. Using this approach, TIM recovers the adiabatic hard-wall and no-interface limits.
We first apply our methodology to tailoring the effective thermal conductivity tensor of a nanomaterial, with potential application in thermal management and routing. Then, we maximize classical size effects while keeping the diffusive transport above a certain threshold, achieving a four-fold improvement with respect to commonly studied staggered configurations. The latter result may have an impact on thermoelectric materials, as explained in the previous section. 

While we have employed a single-MFP model, the developed methodology, along with the interpolation material models, can be readily applied to more sophisticated versions of the BTE. Possible future directions include using the recently-developed anisotropic MFP-BTE~\cite{romano2021efficient}; such an approach would allow modeling a real material using first-principles calculations while taking into account the interplay of phonon-focusing effects and, for example, the possible channels arising during optimization. Another possible extension includes optimizing thermal transport in 2D materials described by the full-scattering operator~\cite{chiloyan2021green,romano2020phonon}.

\backmatter

\section{Acknowledgement}

This work was partially supported by MIT-IBM Watson AI
Laboratory (Challenge No. 2415)

\section{Conflict of interest}

The authors declare that they have no conflict of interest.

\section{Replication of results}

The code developed for this work will be made available as free/open-source software in the next release of OpenBTE~\cite{OpenBTE}.

\begin{appendices}

\section{The Fourier Solver}\label{sec:fourier_solver}

Similarly to previous studies on topology optimization for macroscopic heat conduction to~\cite{gersborg2006topology,evgrafov2011convergence}, we discretize Fourier's law using the finite-volume method (FVM). Material interpolation can be obtained using a space-dependent thermal conductivity,
\begin{equation}\label{eq:fourier}
    \nabla \cdot \left[\rho(x,y) \nabla \g(x,y) \right] = 0.
\end{equation}
We point out that for numerical reasons, we regularized such expression using a small value, $\delta$. However, for clarity, we omit it throughout the text. Once Eq.~\eqref{eq:fourier} is solved, the normalized effective thermal conductivity is evaluated as
\begin{equation}\label{eq:kappa_fourier}
    \bar{\kappa}^{\mathrm{F}}_{xx}= -\int_{-L/2}^{L/2} \bar{\rho}(L/2,y)\nabla \g(L/2,y) \cdot \mathbf{\hat{x}} dy.
\end{equation}
To discretize Eqs.~\eqref{eq:fourier}-\eqref{eq:kappa_fourier}\, we conveniently define the following quantities:
\begin{eqnarray}\label{definitions}
\mathbf{\hat{n}}_{cc'} &=&  \begin{cases}
  \text{normal pointing to } c',& \text{if } c \text{ and } c' \text{ are adjacent (including periodicity)}\\
    0,              & \text{otherwise},
\end{cases}\nonumber \\
K_{cc'} &=&\begin{cases}
  1 ,& \text{if } c \in \textrm{ left and } c' \in \textrm{ right and }  c \textrm{ is adjacent to } c',\\
    0,              & \text{otherwise},
\end{cases}
\end{eqnarray}
where $\mathbf{\hat{n}}_{cc'}$ and $\mathbf{K}$ describe connectivity and external perturbation, respectively. We further define $\mathbf{H} = \left(\mathbf{K}-\mathbf{K}^T\right)$. We discretize Eq.~\eqref{eq:fourier} using the finite-volume method, with the material grid being the same as the discretization grid. Upon integrating Eq.~\eqref{eq:fourier} over the control volume $c$, we have
\begin{eqnarray}\label{eq:assemble}
    \sum_k&&\int_{\partial A_c}\bar{\rho}(x,y)\frac{\partial \g(x,y)}{\partial x_k} n_k dl =\nonumber \\
   =\frac{L}{\sqrt{N}} \mathbf{\bar{J}}_{cc'}\cdot\mathbf{\hat{n}}_{cc'} &=& \frac{L}{\sqrt{N}}\sum_{c'} u_{cc'} \left(\g_{c'} + H_{cc'} - \g_{c}  \right),
\end{eqnarray}
where $\mathbf{\bar{J}}_{cc'}$ is the normalized interfacial thermal flux. To determine $u_{cc'}$ we first write the balance equation at the interface between volume $c$ and $c'$,
\begin{equation}
   \frac{L}{\sqrt{N}} \mathbf{\bar{J}}_{cc'}\cdot\mathbf{\hat{n}}_{cc'}=-2\bar{\rho}_c \left(\g_b-\g_c\right) = -2 \bar{\rho}_{c'}\left(\g_{c'} - \g_b\right), 
\end{equation}
where $\g_b$ is the temperature at the boundary, shared among both volumes (we assume no thermal boundary resistance.) After solving for $\g_b$ (and, for simplicity, assuming we are at an internal volume), we have $\left(L/\sqrt{N}\right)\mathbf{\bar{J}}_{cc'}\cdot \mathbf{\hat{n}}_{cc'} = u_{cc'}\left(\g_{c}- \g_{c'}\right)$, where
\begin{equation}\label{eq:harmonic}
    u_{cc'} = 2\frac{\bar{\rho}_c\bar{\rho}_{c'}}{\bar{\rho}_c + \bar{\rho}_{c'}} \mid \mathbf{\hat{n}}_{cc'}\mid.
\end{equation}
In passing, we point out that Eq.~\eqref{eq:harmonic} is an harmonic average, an approach that has been compared favourably against the arithmetic average, in terms of ability of preventing checkerboard patterns~\cite{gersborg2006topology}. In practice, we use a slightly modified version of Eq.~\ref{eq:harmonic}, $\bar{\rho}_{cc'} = u_{cc'}^\gamma $, where $\gamma$ is a tuning parameter. Note that $\bar{\rho}_{cc'} = \bar{\rho}_{c'c}$. Lastly, Eq.~\eqref{eq:assemble} translates into the linear system
 \begin{equation}\label{eq:linearfourier}
     \sum_{c'} A_{cc'} \g_{c'} = b_c,
 \end{equation}
 where
\begin{eqnarray}\label{lin}
    A_{cc'} &=& \sum_{c''} \bar{\rho}_
    {cc''} \left(\delta_{cc'} -\delta_{c'c''}\right),\nonumber \\
    b_c &=& \sum_{c''}\bar{\rho}_{cc''} H_{cc''}.
\end{eqnarray}
Once Eq.~\ref{lin} is solved, the effective thermal conductivity is evaluated by
\begin{eqnarray}
    \bar{\kappa}^{\mathrm{F}} &=&  \sum_{cc'}\bar{\rho}_{cc'}K_{cc'}\left[\g_c +  1  - \g_{c'}\right]\nonumber \\
    &=& \sum_{cc'} \bar{\rho}_{cc'}K_{cc'} - \sum_c \g_c b_c.
\end{eqnarray}
Depending on the size of $\mathbf{A}$, we solve Eq.~\eqref{eq:linearfourier} either using LU decomposition or an iterative solver; in this last case, the operator associated to Eq.~\eqref{eq:linearfourier} is
\begin{equation}
    \left[\mathcal{L}(\mathbf{x})\right]_c = \sum_{c'}\bar{\rho}_{cc'}\left(x_c -x_{c'}\right),
\end{equation}
and, similarly to the BTE case, the termination criteria is based on the error on $\bar{\kappa}^{\mathrm{F}}$ and $\mid \nabla_{\bar {\boldsymbol \rho}}\bar{\kappa}^{\mathrm{F}}\mid$.

\subsection{Gradient of the Fourier solver}\label{gradient_fourier}

Computing the gradient of $\bar{\kappa}$ with respect to the design field translates into the following chained calculations
\begin{equation}\label{diff2}
    \frac{d \bar{\kappa}^{\mathrm{F}}}{d \bar{\rho}_p}=\frac{\partial \bar{\kappa}^{\mathrm{F}}}{\partial \bar{\rho}_p}+\sum_c\frac{\partial \bar{\kappa}^{\mathrm{F}}}{\partial \g_c}\frac{\partial \g_c}{\partial \bar{\rho}_p}.
\end{equation}
We employ the adjoint method \cite{strang2007computational}, i.e. we differentiate analytically Eq.~\eqref{eq:linearfourier} and then invert it, obtaining
\begin{equation}\label{eq:grad_fourier}
\frac{\partial \bar{\kappa}^{\mathrm{F}}}{\partial \bar{\rho}_p} =  \frac{\partial \bar{\kappa}^{\mathrm{F}}}{\partial \bar{\rho}_p}+\sum_c \g_c^{\mathrm{adj}} C_{cp}
\end{equation}
where
\begin{equation}\label{C3}
    C_{cp} = \sum_{c'}\left[\frac{\partial b_{c'}}{\partial \bar{\rho}_p} \delta_{cc'}-\frac{\partial A_{cc'}}{\partial \bar{\rho}_p}\g_{c'}\right],
\end{equation}
and $\boldsymbol{\g}^{\mathrm{adj}}$ being the solution of the adjoint problem
\begin{equation}\label{dd}
   \sum_{c'} A_{c'c}\g_{c'}^{\mathrm{adj}} = \frac{\partial \bar{\kappa}^{\mathrm{F}}}{\partial \g_c} = -b_c.
\end{equation}
Since $\mathbf{A}$ is symmetric, 
\begin{equation}\label{eq:fourier_symmetry}
     \g_{c}^{\mathrm{adj}} = - \g_c,
\end{equation}
that is, the adjoint solution can be straightforwardly computed using the forward one. A similar result has also been obtained in the context of asymptotic inverse homogeinization~\cite{zhou2008computational}.

To evaluate Eq.~\ref{eq:grad_fourier}, we first note that
\begin{equation}
    \frac{\partial \bar{\rho}_{cc'}}{\partial \bar{\rho}_p} = \gamma\bar{\rho}_{cc'}^{\gamma-1}\left(r_{cc'}\delta_{c p} + r_{c'c}\delta_{c'p}\right),
\end{equation}
where
\begin{equation}
    r_{cc'} =\frac{1}{2}\left( \frac{\bar{\rho}_{cc'}}{\bar{\rho}_c}\right)^2. 
\end{equation}
Then, after some algebra, we have
\begin{eqnarray}\label{final}
\sum_c \g_c^{\mathrm{adj}}\frac{\partial b_c}{\partial \bar{\rho}_p} &=& \sum_c r_{pc} \left(H_{pc} \g_p^{\mathrm{adj}} + H_{cp} \g_c^{\mathrm{adj}} \right) \\
-\sum_{cc'}\g_c^{\mathrm{adj}}\frac{\partial A_{cc'}}{\partial \bar{\rho}_p}\g_{c'} &=& \sum_c r_{pc}\bigg(\g_c \g_p^{\mathrm{adj}}   + \g_p  \g_c^{\mathrm{adj}} -\g_p \g_p^{\mathrm{adj}}- \g_c\g_c^{\mathrm{adj}}  \bigg),\nonumber\\
\frac{\partial \bar{\kappa}^{\mathrm{F}}}{\partial \bar{\rho}_p} &=& \sum_c r_{pc}\left(K_{pc} + K_{cp} + H_{pc} \g_c -  H_{pc}\g_p\right).
\end{eqnarray}
Putting everything together, we have
\begin{eqnarray}\label{eq:tmp2}
    \frac{d \bar{\kappa}^{\mathrm{F}}}{d \bar{\rho}_p} &=& \sum_c r_{pc}\Bigg[H_{pc} \g_p^{\mathrm{adj}} + H_{cp} \g_c^{\mathrm{adj}} +\g_c \g_p^{\mathrm{adj}}   +  \g_p \g_c^{\mathrm{adj}} -\g_p \g_p^{\mathrm{adj}}-\nonumber \\ &-& \g_c\g_c^{\mathrm{adj}} +K_{pc} + K_{cp} + H_{pc} \g_c -  H_{pc}\g_p \Bigg].
\end{eqnarray}
Lastly, using Eq.~\ref{eq:fourier_symmetry} in Eq.~\ref{eq:tmp2}, we have
\begin{eqnarray}
    \frac{d \bar{\kappa}^{\mathrm{F}}}{d \bar{\rho}_p} = \sum_c r_{pc}\Bigg[ \left(\g_p - \g_c\right)^2  -2\left(\g_p - \g_c \right)\left(K_{pc}- K_{cp} \right) +\nonumber \\ 
    + K_{pc} + K_{cp}  \Bigg],
\end{eqnarray}
which is the actual expression implemented.

\section{The BTE solver}\label{sec:bte_solver}

Several deterministic approaches have been developed to solve the BTE in arbitrary structures, including the lattice Boltzmann method~\cite{nabovati2011lattice}, spherical harmonics~\cite{mittal2011hybrid} and finite-volume methods~\cite{murthy2005review,romano2011multiscale}. In this work, we adopt the latter approach, where both the real- and angular-space are integrated over a control volume. For simplicity and with no loss of generality, we discretize the angular space uniformly. Specifically, we choose $M=48$ angular bins, for which $\bar{\kappa}$ converges within $<1\%$ error for both regular and random structures, and for all the grid resolutions considered in this work. We integrate both sides of Eq.~\eqref{eq:bte_gray} over the control angle $\Delta \phi$ centered at $\phi_\mu$. Assuming that the unknowns are constant within the single angular cell, we have
\begin{eqnarray}\label{eq:integro}
  \Lambda \mathbf{S}_\mu\cdot\nabla \g_\mu(\mathbf{r}) +  \g_\mu(\mathbf{r}) = \frac{1}{M}\sum_{\mu'} \g_{\mu'}(\mathbf{r}),
\end{eqnarray}
where 
\begin{equation}\label{eq:s_discr}
\mathbf{S}_\mu = \frac{1}{\Delta \phi}\int_{\phi_\mu-\Delta\phi/2}^{\phi_\mu+\Delta\phi/2} \s d\phi = \mathrm{sinc}\left(\frac{\Delta \phi}{2}\right)\mathbf{\hat{s}}(\phi_\mu).
\end{equation}
In Eq.~\eqref{eq:s_discr}, we use the notation $\mathrm{sinc}(x) = \sin(x)/x$. The spatial discretization is carried out using the upwind, finite-volume scheme~\cite{murthy2005review,romano2011multiscale}. Averaging Eq.~\eqref{eq:integro} over the control volume $c$ and applying Gauss' law gives
\begin{equation}\label{eq:semi_bte}
  \kn \sum_{c'} \g_{\mu}(\mathbf{r}_{cc'}) \mathbf{S}_\mu \cdot \mathbf{\hat{n}}_{cc'} +  \g_\mu(\mathbf{r}_c) = \frac{1}{M}\sum_{\mu'} \g_{\mu'}(\mathbf{r}_{c}),
\end{equation}
where $\kn =   \Lambda\frac{\sqrt{N}}{L}$ is defined as the Knudsen number, $\mathbf{r}_c$ is the centroid of volumes $c$, and $\mathbf{r}_{cc'}$ is the centroid of the face between volume $c$ and $c'$. The term $\g_\mu(\mathbf{r}_{cc'})$ is evaluated using Eq.~\eqref{eq:bc} and upwind differentiation,
\begin{eqnarray}\label{eq:upwind}
   \g_\mu(\mathbf{r}_{cc'})\mathbf{S}_\mu\cdot &\mathbf{\hat{n}}_{cc'}&\approx  \g_{\mu c} (\mathbf{S}_\mu\cdot \mathbf{\hat{n}}_{cc'})^+ + \nonumber \\
   &+&(\mathbf{S}_\mu\cdot \mathbf{\hat{n}}_{cc'})^-\biggl[t_{cc'} \left(\g_{\mu c'}+ H_{cc'} \right) 
   +\left(1-t_{cc'}\right)\g_{cc'}^B\biggr],
\end{eqnarray}
where $(x)^+$ is $\mathrm{max}(0,x)$, $(x)^- = \mathrm{min}(0,x)$, $t_{cc'}$ is defined in Eq.~\eqref{eq:transmission}, and $H_{cc'}$ is introduced in Sec.~\ref{sec:fourier_solver}. In Eq.~\eqref{eq:upwind}, $\g_{cc'}^B$ is obtained by discretizing Eq.~\eqref{eq:TB},
\begin{equation}\label{eq:tb_discretized}
     \tilde{T}^B_{cc'}=\frac{1}{2}\sum_\mu \g_{\mu c} (\mathbf{S}_\mu \cdot \mathbf{\hat{n}}_{cc'})^+.
\end{equation}
Combining Eqs.~\eqref{eq:semi_bte},~\eqref{eq:upwind} and~\eqref{eq:tb_discretized}, we have the following \emph{matrix} linear system
\begin{equation}\label{mm2}
 \sum_{\mu'c'}G_{\mu c}^{\mu'c'}\g_{\mu'c'}= P_{\mu c},
\end{equation}
with $c=1,...,N-1, \mu =1,...,M-1$, and
\begin{equation}\label{matrix}
    G_{\mu c}^{\mu'c'} = \delta_{cc'}\delta_{\mu\mu'}\left(1+D_{\mu'c'}\right) + \delta_{\mu\mu'} A_{\mu cc'} + \delta_{cc'}\left(B_{c\mu\mu'}-\frac{1}{M}\right).
\end{equation}
The notation $\delta_{cc'}$ refers to the Kronecker delta. The terms in Eq.~\eqref{matrix} are
\begin{eqnarray}
D_{\mu' c'} &=& \sum_{c''} g^+_{\mu' c'c''}, \nonumber \\
A_{\mu cc'} &=&  t_{cc'}g^-_{\mu cc'},\nonumber \\
P_{\mu c}&=&-\sum_{c'} H_{cc'} t_{cc'} g_{\mu cc'}^{-}\nonumber \\
B_{\mu \mu' c} &=&  \sum_{c''}  h_{\mu c}^{\mu'c''}\left(1-t_{cc''}\right).
\end{eqnarray}
where
\begin{eqnarray}\label{eq:def}
    g_{\mu cc'}^{+(-)} &=&\mathrm{Kn} (\mathbf{S}_\mu \cdot \mathbf{\hat{n}}_{cc'})^{+(-)}, \nonumber \\
    h_{\mu c}^{\mu'c''} &=&  g_{\mu cc''}^{-}g_{\mu' cc''}^{+} \alpha_{cc''}^{-1}.
\end{eqnarray}
In Eq.~\eqref{eq:def}, we defined $\alpha_{cc''} = \sum_{\mu''}g_{\mu''cc''}^+$. Lastly, the effective thermal conductivity is computed by discretizing Eq.~\eqref{eq:kappa_bte} using a similar procedure, yielding
\begin{eqnarray}
    \bar{\kappa} &=& -\frac{1}{f}\sum_{\mu cc'} t_{cc'}K_{cc'} [\g_{\mu c} g^+_{\mu cc'}  +   \left(\g_{\mu c'}+1\right) g^-_{\mu cc'}] =\nonumber \\&=& -\frac{1}{f}\sum_{\mu cc'} t_{cc'} K_{cc'}g_{\mu cc'}^- + \frac{1}{f}\sum_{\mu c} P^{\mathrm{adj}}_{\mu c} \g_{\mu c},
\end{eqnarray}
where $f = \mathrm{Kn}^2 M/2$ and
\begin{equation}
 P^{\mathrm{adj}}_{\mu c} = -\sum_{c'} t_{cc'} H_{cc'}g_{\mu cc'}^+.
\end{equation}
Equation~\eqref{matrix} can be cast into a standard linear system $\mathbf{A}\mathbf{x}=\mathbf{b}$, with $A\in\mathbb{R}^{NM,NM}$ and $\mathbf{b}\in\mathbb{R}^{NM}$. In practice, we use a Krylov-subspace, matrix-free approach (specifically, LGMRES~\cite{baker2005technique}) thus $\mathbf{A}$ is never evaluated. To this end, we first implement the linear operator 
\begin{equation}
\mathbf{X} \rightarrow \mathcal{L}(\mathbf{X}),
\end{equation}
with $\mathcal{L}:\mathbb{R}^{NM,NM}\rightarrow \mathbb{R}^{NM,NM}$, defined as
\begin{equation}\label{eq:operator}
    \left[\mathcal{L}(\mathbf{X})\right]_c =  D_{\mu c} X_{\mu c} + \sum_{c'} A_{\mu cc'}  X_{\mu c'} + \sum_{\mu '} \left(B_{c\mu\mu'}-\frac{1}{M}\right) X_{\mu' c}; 
\end{equation}
then, we solve the linear system
\begin{equation}
    \bar{\mathcal{L}}(\mathbf{x})=\mathbf{b},
\end{equation}
where
\begin{eqnarray}\label{flattening}
     \bar{\mathcal{L}} &=&   \mathcal{F}\circ \mathcal{L}\circ \mathcal{F}^{-1}:\mathbb{R}^{NM}\rightarrow \mathbb{R}^{NM} \nonumber \\
     \mathbf{b} &=& \mathcal{F}(\mathbf{P}).
\end{eqnarray}
The terms $\mathcal{F}:\mathbb{R}^{N,M}\rightarrow \mathbb{R}^{NM}$ and $\mathcal{F}^{-1}:\mathbb{R}^{NM}\rightarrow \mathbb{R}^{N,M}$ are the flattening and reshaping operator, respectively. Lastly, as a first guess to the LGMRES solver, we use the solution from the previous optimization iteration. As a result, we gain a reduction in the number of operator calls of about $30\%$, $50\%$ and, $70\%$ for grids with $N = 20\times20$, $40\times40$ and $60\times60$, respectively (based on an exemplary optimization with 50 iterations, $\beta = 2$, and $M = 48$).

\subsection{Gradient of the BTE solver}

In this section, we detail on the calculation of gradient of $\bar{\kappa}$ with respect to the projected density $\bar{\rho}$. To this end, we conveniently defined (for a reason that will be apparent later) the \emph{scaled} effective thermal conductivity $\bar{\kappa}^s = f\bar{\kappa}$, with $f$ defined in the previous section. Thus, our goal is to evaluate
\begin{eqnarray}\label{eq:grad2}
   \frac{d \bar{\kappa}^s}{d \bar{\rho}_p} &=& \sum_{\mu c}\frac{\partial \bar{\kappa}^s}{\partial \g_{\mu c}}\frac{\partial \g_{\mu c}}{\partial   \bar{\rho}_p}+ \frac{\partial \bar{\kappa}^s}{\partial \bar{\rho}_p}.
\end{eqnarray}
The calculation of Eq.~\eqref{eq:grad2} is aided by the adjoint method~\cite{strang2007computational}, which, for our matrix linear system, reads
\begin{eqnarray}\label{eq:grad3}
   \frac{d \bar{\kappa}^s}{d \bar{\rho}_p} =\sum_{\mu c} \g_{\mu c}^{\mathrm{adj}} C_{\mu cp} + \frac{\partial \bar{\kappa}^s}{\partial \bar{\rho}_p},
\end{eqnarray}
where $\mathbf{\g}^{\mathrm{adj}}$ is the solution of the matrix linear system
\begin{equation}\label{eq:reverse}
    \sum_{\mu'c'}  G_{\mu' c' }^{\mu c}\g_{\mu'c'}^{\mathrm{adj}} = \frac{\partial \bar{\kappa}^s}{\partial \g_{\mu c}} = P^{\mathrm{adj}}_{\mu c} ,
\end{equation}
and
\begin{equation}\label{eq:cip}
    C_{\mu c p} = \sum_{\mu'c'}\left(\frac{\partial P_{\mu'c'}}{\partial \bar{\rho}_p} \delta_{cc'}\delta_{\mu\mu'}- \g_{\mu'c'}\frac{\partial G_{\mu c}^{\mu'c'}}{\partial \bar{\rho}_p} \right).
\end{equation}
Motivated by the relationship between the forward and adjoint solution of the Fourier solver [Eq.~\eqref{eq:fourier_symmetry}], we seek a similar link between $\mathbf{\g}$ and $\mathbf{\g}^{\mathrm{adj}}$. To this end, we first compare the forward and adjoint solvers
\begin{eqnarray}
\sum_{\mu' c'}\left[U_{\mu c}^{\mu' c'} + F_{\mu c}^{\mu' c'}   \right]\g_{\mu'c'} &=& -\sum_{c'} t_{cc'}  H_{cc'} g_{\mu cc'}^{-}\nonumber \\
\sum_{\mu' c'}\left[U_{\mu' c'}^{\mu c} + F_{\mu' c'}^{\mu c}   \right] \g_{\mu'c'}^{\mathrm{adj}} &=& -\sum_{c'}t_{cc'}H_{cc'} g_{\mu cc'}^+,
\end{eqnarray}
where
\begin{eqnarray}
    U_{\mu c}^{\mu ' c'} &=& \delta_{cc'}\delta_{\mu\mu'}\left(1+D_{\mu c}\right) + \delta_{\mu\mu'} A_{\mu cc'} -\delta_{cc'}\frac{1}{M}\nonumber \\
     F_{\mu c}^{\mu ' c'} &=&  \delta_{cc'}\sum_{c''}\alpha_{cc''}^{-1}  g_{\mu cc''}^{-}g_{\mu' cc''}^+\left(1-t_{cc''}\right).
\end{eqnarray}
We make the change of variables $\mu \rightarrow -\mu$ and $\mu' \rightarrow -\mu'$ in Eq.~\eqref{eq:reverse}, where $\mathbf{\hat{s}}(\phi_{-\mu}) = -\mathbf{\hat{s}}(\phi_{\mu})$. Correspondingly, we note the following equalities
\begin{eqnarray}\label{eq:equalities}
    g_{\mu cc'}^+ &=& -g_{\mu c'c}^- \nonumber \\
    g_{\mu cc'}^- &=& -g_{\mu c'c}^+ \nonumber \\
    g_{\mu cc'}^+ &=& -g_{-\mu cc'}^-\nonumber \\
    g_{\mu cc'}^- &=& -g_{-\mu cc'}^+.
\end{eqnarray}
Combining the first and the third property, we have $g_{-\mu cc'}^- =g_{\mu c'c}^-$, which translates into $A_{-\mu cc'} =A_{\mu c'c} $. Furthermore, $\delta_{-\mu-\mu'} = \delta_{\mu\mu'}$, and
\begin{equation}\label{eq:D}
D_{-\mu c} = \sum_{c''} g_{-\mu cc''}^+ = -\sum_{c''} g_{-\mu cc''}^- = \sum_{c''} g_{\mu c''c}^+ = D_{\mu c},
\end{equation}
where we use $g_{-\mu cc''}^+ = g_{\mu c''c}^+$. The last relationship of Eq.~\eqref{eq:D} can be proved noting that
\begin{equation}
\sum_{c''}\left[(\mathbf{S}_\mu\cdot \mathbf{\hat{n}}_{cc''})^+ - (\mathbf{S}_\mu\cdot \mathbf{\hat{n}}_{c''c})^+\right] = \mathbf{S}_\mu \cdot \sum_{c''} \mathbf{n}_{cc''} = 0,
\end{equation}
i.e. the sum of the normal to the sides of a square, pointing outward, is zero. Lastly,
\begin{equation}\label{eq:Psym}
 P_{-\mu c}^{\mathrm{adj}}= -\sum_{c'}t_{cc'}H_{cc'} g_{-\mu cc'}^+ =    \sum_{c'}t_{cc'}H_{cc'} g_{\mu cc'}^- = -P_{\mu c}.
\end{equation}
Combining these equations leads to
\begin{eqnarray}
  U_{-\mu' c'}^{-\mu c} &=&   U_{\mu c}^{\mu' c'}\nonumber \\
    F_{-\mu' c'}^{-\mu c} &=& F_{\mu c}^{\mu' c'}\\ \nonumber
    G_{-\mu' c'}^{-\mu c} &=& G_{\mu c}^{\mu' c'}
\end{eqnarray}
Hence, Eq.~\eqref{eq:reverse} becomes
\begin{equation}
    \sum_{\mu' c'} G_{\mu c}^{\mu' c'}  \left(-\g_{-\mu'c'}^{\mathrm{adj}}\right) = P_{\mu c},
\end{equation}
from which we deduce that
\begin{equation}\label{eq:bte_symmetry}
    \g_{\mu c}^{\mathrm{adj}} = -\g_{-\mu c}.
\end{equation}
In light of this result, we can therefore compute the adjoint solution directly from the forward one, without solving a linear system again. Note that by scaling $\bar{\kappa}$ we were able to have the forward and the adjoint solution dimensionally consistent.

Let's now evaluate Eq.~\eqref{eq:grad2}. We begin by noting that
 \begin{equation}
\frac{\partial t_{cc'}}{\partial \bar{\rho}_p} = r_{cc'} \delta_{cp} + r_{c'c} \delta_{c'p},
\end{equation}
where $r_{cc'}=1/2\left(t_{cc'}/\bar{\rho}_c \right)^2$. We expand the terms appearing in Eqs.~\eqref{eq:grad3}-\eqref{eq:cip},
\begin{eqnarray}
\frac{\partial \bar{\kappa}^s }{\partial \bar{\rho_p}} &=& -\sum_{\mu c} r_{pc}\left(K_{pc}g_{\mu pc}^- + K_{cp} g_{\mu cp}^- +g_{\mu cp}^+ H_{cp}\g_{\mu c} + g_{\mu pc}^+ H_{pc}\g_{\mu p} \right) \nonumber,\\
\sum_{\mu c}\g^{\mathrm{adj}}_{\mu c}\frac{\partial P_{\mu c}}{\partial \bar{\rho}_p}&=&\sum_{\mu c} r_{pc}H_{pc}\left(g_{\mu pc}^- \g_{-\mu p} -g_{\mu cp}^- \g_{-\mu c}\right)\nonumber,\\
-\sum_{\mu \mu' c c'}\g_{\mu c}^{\mathrm{adj}}\frac{\partial G_{\mu c}^{\mu' c'}}{\partial \bar{\rho}_p}\g_{\mu' c'} &=& \sum_{\mu c} r_{pc} \Biggl[T_{-\mu p}T_{\mu c}g_{\mu pc}^- + T_{-\mu c}T_{\mu p}g_{\mu cp}^- - \nonumber \\ &-&\sum_{\mu'}\left(\g_{-\mu c}\g_{\mu' c}h_{\mu c}^{\mu' p}+\g_{-\mu p}\g_{\mu' p}h_{\mu p}^{\mu' c} \right)\biggr] 
\end{eqnarray}
where we used Eq.~\eqref{eq:bte_symmetry}. Finally, putting everything together, we have
\begin{eqnarray}
\frac{\partial \bar{\kappa}^s}{\partial \bar{\rho}_p} &=& \sum_{\mu c}r_{pc}\biggl[H_{pc}\left(g_{\mu pc}^- \g_{-\mu p} -g_{\mu cp}^- \g_{-\mu c}-g_{\mu pc}^+ \g_{\mu p} + g_{\mu cp}^+ \g_{\mu c}\right)- \nonumber\\
&-&K_{pc}g_{\mu pc}^- - K_{cp} g_{\mu cp}^- +T_{-\mu p}T_{\mu c}g_{\mu pc}^- + T_{-\mu c}T_{\mu p}g_{\mu cp}^- -\nonumber \\ &-&\sum_{\mu'}\left(\g_{-\mu c}\g_{\mu' c}h_{\mu c}^{\mu' p}+\g_{-\mu p}\g_{\mu' p}h_{\mu p}^{\mu' c}\right)
\biggr].\end{eqnarray}
Lastly, we note that thanks to Eq.~\eqref{eq:Psym}, we can compute $\bar{\kappa}^s$ using directly $\mathbf{P}$, i.e.
\begin{equation}
\bar{\kappa}^s = \sum_{\mu c}P_{\mu c} \g_{-\mu c} - \sum_{\mu cc'} t_{cc'}K_{cc'}g_{\mu cc'}^-,
\end{equation}
thus sparing the computation of $\mathbf{P}^{\mathrm{adj}}$.

\end{appendices}


\bibliography{biblio}% common bib file


\end{document}